\documentclass{appolb}
\usepackage{epsfig}
\usepackage[T1]{fontenc}
\usepackage{bm}
\def\Lmunu{L_{\mu\nu}}
\def\Wmunu{W^{\mu\nu}}
\def\openone{\leavevmode\hbox{\small1\kern-3.8pt\normalsize1}}
\def\beq{\begin{equation}}
\def\eeq{\end{equation}}
\def\be{\begin{eqnarray}}
\def\ee{\end{eqnarray}}
\def\lsim{\buildrel < \over {_{\sim}}}

\begin{document}
\eqsec  
\title{How much nuclear physics do we need, to understand \\
the neutrino nucleus cross section ? %
\thanks{Presented at the 45th Winter School in Theoretical Physics ``Neutrino Interactions: from Theory to Monte Carlo Simulations'', L\k{a}dek-Zdr\'oj, Poland, February 2--11, 2009.}%
}
\author{Omar Benhar
\address{INFN and Dipartimento di Fisica, ``Sapienza'' Universit\`a di Roma. \\ I-00185 Roma, Italy.}
}
\maketitle
\begin{abstract}
Over the past two decades, electron scattering experiments have clearly exposed the 
limits of the independent particle model description of atomic nuclei.
I will briefly outline the dynamics leading to the appearance of strong correlation 
effects, and their impact on the electroweak nuclear cross sections in the impulse 
approximation regime.
\end{abstract}

\section{Introduction}
\label{intro}

The theoretical description of nuclear structure and dynamics involves severe 
difficulties, arising from both the nature of strong interactions and the complexity 
of the quantum mechanical many-body problem.
 
In the absence of {\em ab initio} approaches, one has to resort to nuclear models,  
 based on effective degrees of freedom, protons and neutrons, and 
phenomenological effective interactions. 
The avaliable empirical information shows that the nucleon-nucleon (NN)
potential exhibits a rich operatorial structure, including spin-isospin
dependent and non central components. 

Due to the complicated nuclear hamiltonian, the exact solution of the 
many body Schr\"odinger equation 
turns out to be a highly challenging computational task. On the other 
hand, nuclear systematics suggests that important features of nuclear dynamics
can be described using the independent particle model, based on the replacement of 
the NN potential with a {\em mean field}. This is in fact the main tenet of the nuclear 
shell model, which proved exceedingly successful in describing a variety of nuclear 
properties. 

The simplest implementation of the independent particle picture is the 
 Fermi gas (FG) model, in which the nucleus is seen as a degenerate Fermi 
gas of neutrons and protons, bound with constant 
energy. 

In spite of all the accomplishemnts of the shell model, it has to be kept in 
mind that in their classic 
nuclear physics book, first published in 1952, 
Blatt and Weisskopf warn the reader that ``the limitation of any independent 
particle model lies in its inability to encompass the correlation between 
the positions and spins of the various particles in the system'' \cite{BW}.

In recent years, electron scattering experiments have provided overwhelming 
evidence of correlations in nuclei, whose description requires the 
use of realistic NN potentials within the formalism of nuclear many-body 
theory.

In this lectures, after briefly recalling few basic facts on nuclear dynamics
beyond the independent particle model, I will discuss the impact of 
correlation effects on the electroweak nuclear cross sections in the impulse
approximation regime.

\section{Basic facts on nuclear structure and dynamics}
\label{facts}

One of the most distinctive features of the NN interaction 
can be inferred from the analysis of the nuclear charge distributions, measured by 
elastic electron-nucleus scattering experiments.

\begin{figure}[hbt]
\centerline%
{\psfig{figure=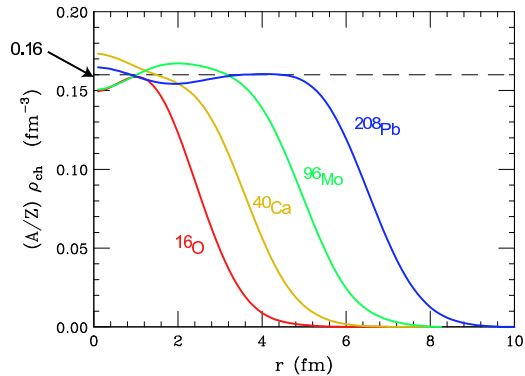,angle=00,width=6.8cm}}
\caption{\small Radial dependence of the charge density distributions of
different nuclei. \label{nucdens}}
\end{figure}

As shown in Fig. \ref{nucdens}, the densities of different nuclei, normalized to 
the number of protons, exhibit {\it saturation}, their 
value in the nuclear interior ($\rho_0 \sim 0.16$ fm$^{-3}$) being nearly constant 
and independent of the mass number $A$. 
This observation tells us that nucleons cannot be packed together too tightly,
thus pointing to the existence of NN {\em correlations} in coordinate space.

Correlations affect 
the {\em joint} probability of finding two nucleons at positions 
${\bf x}$ and ${\bf y}$, usually written in the form
\beq
\rho({\bf x},{\bf y}) = \rho({\bf x})\rho({\bf y}) g({\bf x},{\bf y}) \ ,
\label{repuls1}
\eeq
where $\rho({\bf x})$ is the probability of finding a 
nucleon at position ${\bf x}$.  In the absence of correlations $g({\bf x},{\bf y}) = 1$. On the 
other hand, saturation of nuclear densities indicates that 
\beq
|{\bf x}-{\bf y}| \lsim r_c \ \Longrightarrow  \ g({\bf x},{\bf y}) \ll 1 \ ,
\label{repuls2}
\eeq
$r_c$ being the correlation range.

\begin{figure}[hbt]
\centerline%
{\psfig{figure=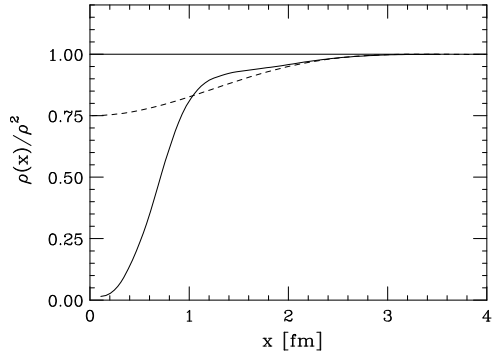,angle=00,width=6.5cm}}
\caption{\small Spin-isospin averaged NN radial correlation function in 
isospin symmetric nuclear matter at uniform density $\rho_0=0.16$ fm$^{-3}$.
The solid line shows the full result of the calculation of Ref. \cite{compg}, 
while the dashed line only includes statistical correlations.
\label{g}}
\end{figure}

Nucleons obey Fermi statistics, and may therefore 
repel one another even in the absence of dynamical interactions. To see this, 
consider a degenerate FG consisting of equal number of
protons and neutrons at uniform density $\rho$. In this case Eq.(\ref{repuls1})
  reduces to
\beq
\rho(|{\bf x}-{\bf y}|) = \rho^2 g_F(|{\bf x}-{\bf y}|) \ ,
\eeq
with the correlation function $g_F(x)$ displayed by the dashed line in Fig. \ref{g}.
It clearly appears that the effects of statistical correlations, while being 
clearly visible, is not too strong. The probability of finding two nucleons at
relative distance $x \ll 1$ fm is still very large.

In the early days of nuclear physics, just after the neutron had been discovered and
the existence of neutron stars had been proposed, Tolman, Oppenheimer and 
Volkoff \cite{T,OV} carried out 
the first studies of the stability of neutron stars, modeled as a gas of 
noninteracting 
particles at zero temperature. Their work was aimed at determining whether the 
degeneracy pressure, resulting from the repulsion induced by Pauli exclusion 
principle, could become strong enough to balance the gravitational pull, thus giving 
rise to a stable star. These calculations led to predict a maximum 
neutron star mass $\sim 0.8$~M$_\odot$, M$_\odot$ being the mass of the sun, 
to be compared to the results of most experimental measurements yelding 
values $\sim 1.4$~M$_\odot$. 
The observation of neutron stars with masses
largely exceeding the upper limit determined in Refs.\cite{T,OV} can be regarded 
as a striking evidence of the failure of the description of nuclear 
systems based on the FG model.
To explain the observed neutron stars masses, the effects of nuclear dynamics have 
to be explicitely taken into account.

The strength of {\em dynamical} NN correlations is illustrated by the solid line
of Fig. \ref{g}, showing the NN radial correlation function in
nuclear matter at uniform density $\rho_0=0.16$ fm$^{-3}$, obtained from 
the variational approach discussed in the Section \ref{NMBT}.
Comparison with the dashed line, computed including statistical correlations only,  
clealry shows that the dynamical effects dominate.

\section{The nucleon-nucleon interaction}
\label{force}

The NN interaction can be best studied in the two-nucleon system.
There is only one NN bound state, the nucleus of deuterium, or
deuteron, consisting of a proton
and a neutron coupled to total spin and isospin $S=1$ and $T=0$, respectively.
This is clear manifestation of the fact that nuclear forces 
are {\em spin-isospin dependent}.

Another important piece of information can be inferred from the
observation that the deuteron exhibits a nonvanishing electric quadrupole moment,
implying that its charge distribution is not spherically symmetryc. Hence,
the NN interaction is {\em noncentral}.

Besides the properties of the two-nucleon bound state, the large data set of phase
shifts measured in NN scattering experiments ($\sim$ 4000 data points, corresponding
to energies up to pion production theshold) provides valuable additional information
on the nature of NN forces.

Back in the 1930s, Yukawa suggested that nuclear interactions were mediated 
by a particle of mass $\sim 100$ MeV, that was later identified 
with the pion. The one pion exchange (OPE) mechanism provides a fairly accurate 
description of the long range behavior of the NN interaction, as it explains the measured 
NN scattering phase shifts in states of high angular momentum. 

At intermediate and short range more complicated processes, involving
the exchange of two or more pions (possibly interacting among themselves) or
heavier particles, like the $\rho$ and $\omega$ mesons, have to be taken
into account. Moreover, when their relative distance becomes very small
($\lsim 0.5$ fm) nucleons, being composite and finite in size,
are expected to overlap. In this regime, NN interactions should in principle
be described in terms of interactions between nucleon
constituents, i.e. quarks and gluons, as dictated by quantum chromodynamics
(QCD), which is believed to be the fundamental theory of strong interactions.

Phenomenological potentials describing the {\em full} NN interaction are generally
written in the form
\beq
v = v_{\pi} + v_{R} \ ,
\label{phenv:1}
\eeq
where $v_{\pi}$ is the OPE potential, while $v_{R}$
describes the interaction at intermediate and short range. 

The spin-isospin dependence
and the noncentral nature of the potential can be properly accounted for
rewriting Eq. (\ref{phenv:1}) in the form
\beq
v_{ij} = \sum_{ST} \left[ v_{TS}(r_{ij}) + \delta_{S1} v_{tT}(r_{ij}) S_{ij} \right]
P_S \Pi_T\ ,
\label{pot:TS}
\eeq
where $S$ and $T$ denote the total spin and isospin of the interacting pair, 
$P_S$ and $\Pi_T$ are the corresponding projection operators and 
\beq
S_{ij} = \frac{3}{r_{ij}^2}\ 
({\bm \sigma}_i \cdot {\bf r}_{ij}) ({\bm \sigma}_j \cdot {\bf r}_{ij})
 - ({\bm \sigma}_i \cdot {\bm \sigma}_j ) \ ,
\eeq
reminiscent of the operator describing the interaction between two magnetic
dipoles, accounts for the presence of non central contributions. 

The functions $v_{TS}(r_{ij})$ and $v_{tT}(r_{ij})$ describe the radial dependence 
of the
interaction in the different spin-isospin channels, and reduce to the corresponding 
components
of the OPE potential at large $r_{ij}$. Their shapes are chosen in 
such a way as to reproduce the available NN data (deuteron binding energy, 
charge radius and quadrupole moment and the NN scattering phase shifts).

\begin{figure}[hbt]
\centerline
{\psfig{figure=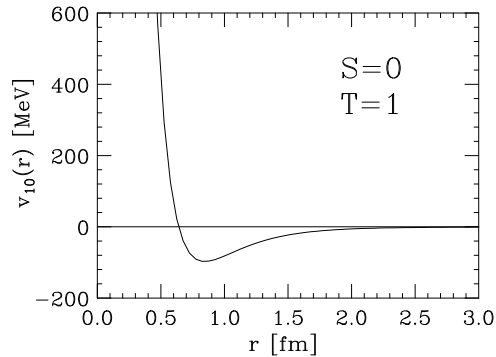,angle=00,width=6.5cm}}
\caption{\small Radial dependence of the NN potential describing the interaction
between two nucleons in the state of total spin and isospin $S = 0$ and $T = 1$.
\label{NN:pot}}
\end{figure}

As an example, Fig. \ref{NN:pot} shows the potential acting beteween two nucleons
with $S=0$ and $T=1$. The presence of the repulsive core inducing strong 
short range correlations (compare to Fig. \ref{g}) is apparent.

Although state-of-the-art parametrizations of the NN potential \cite{WSS} 
have a more complex operatorial structure, including  non static and 
charge symmetry breaking components, the simple form (\ref{pot:TS}) has
the advantage of being easily applicable, and still 
allows one to obtain a reasonable description of the two-nucleon bound 
and scattering states.

\section{Nuclear many body theory}
\label{NMBT}

According to the {\em paradigm} of nuclear many-body theory (NMBT) the nucleus 
can be viewed as a collection of
$A$ pointlike protons and neutrons, whose dynamics are described by the
nonrelativistic hamiltonian
\beq
H = \sum_i \frac{{\bf p}_i^2}{2m} + \sum_{j>i} v_{ij}
+ \sum_{k>j>i} V_{ijk}\ ,
\label{hamiltonian}
\eeq
where ${\bf p}_i$ and $m$ denote the momentum of the $i$-th nucleon and
its mass, respectively.                                                                                         
The determination of the two-body potential $v_{ij}$ has been outlined in the 
previous section. The inclusion of the three-nucleon interaction, whose 
contribution to the energy satisfies 
$\langle~V_{ijk}~\rangle~\ll~\langle~v_{ij}~\rangle$,
is required to account for the binding energy of the three-nucleon 
systems \cite{PPCPW}.

It is very important to realize that in NMBT the dynamics is fully
specified by the properties of {\em exactly solvable}
system, having $A \leq 3$, and does not suffer from the uncertainties involved in
many body calculations. Once the nuclear hamiltonian is fixed, calculations
of nuclear observables for a variety of systems, ranging from the deuteron to
neutron stars, can be carried out without making use of any adjustable
parameters.

The predictive power of the dynamical model based on the  hamiltonian
of Eq.(\ref{hamiltonian}) has been extensively tested by computing the energies 
of the ground and low-lying excited states of nuclei with $A \leq 12$. 
The results of these studies, in which the many body Schr\"odinger 
equation is solved {\em exactly} using stochastic methods,
turn out to be in excellent agreement with experimental data \cite{WP}.

Accurate calculations can also be carried out for uniform nuclear matter, 
exploiting translational invariance and using the stochastic method
\cite{AFDMC}, the variational approach \cite{Akmal98}, or G-matrix perturbation 
theory \cite{Baldo00}. 

In the variational approach, the nuclear states are written in such a
way as to incorporate the correlation structure induced by NN interactions. 
In the case of uniform nuclear matter, they can be obtained from the states
 of the noninteracting FG through the transformation
\beq
|n\rangle = F |n_{FG}\rangle \ ,
\eeq
with $F$ written in the form
\begin{equation}
F=\mathcal{S}\prod_{ij} f_{ij} \  .
\end{equation}
The structure of the two-body correlation operator $f_{ij}$
reflects the complexity of the NN potential, described by Eq.(\ref{pot:TS}),
while the symmetrization operator $\mathcal{S}$ is needed to account for the
fact that $[f_{ij},f_{jk}] \neq 0$.
The shapes of the radial functions $f_{TS}(r_{ij})$ and $f_{tT}(r_{ij})$ are 
determined by functional minimization of the expectation value of the
hamiltonian (\ref{hamiltonian}) in the correlated ground state.

The formalism based on correlated wave functions is ideally suited to 
carry out calculations of nuclear matter properties strongly 
affected by correlation effetcs.

The hole spectral function $P_h({\bf k},E)$, yielding the probability
of removing a nucleon of momentum ${\bf k}$ from the nuclear ground 
state leaving the residual system with excitation energy $E$ \cite{pke},
can be written in the form 
\beq
P_h({\bf k},E) =  \frac{1}{\pi}
\frac{ Z_k^2 \ {\rm Im} \Sigma({\bf k},\epsilon_k) }
{ (E + \epsilon_k)^2 +
          [Z_k {\rm Im}\ \Sigma({\bf k},\epsilon_k)]^2 } 
 + P^B_h({\bf k},E) \ ,
\label{Phke}
\eeq
with $\epsilon_k$ defined by the equation
\beq
\epsilon_k = \epsilon^0_k + {\rm Re}\ \Sigma({\bf k},\epsilon_k) \ ,
\eeq
where $\epsilon^0_k = |{\bf k}|^2/2m$ and 
$\Sigma({\bf k},E)$ is the nucleon self energy.

The first term in the right hand side of equation (\ref{Phke}) describes the 
spectrum of a system of independent {\em quasiparticles} of momentum 
$|{\bf k}|<k_F$, $k_F$ being the Fermi momentum, moving in
a complex mean field whose real and imaginary parts determine the quasiparticle
effective mass and lifetime, respectively. In the FG model
this term shrinks to a $\delta$-function and $Z_k=1$.
The presence of the second term is a 
pure correlation effect. In the FG model $P^B_h({\bf k},E) = 0$, while 
in the presence of interactions the correlation term is the only one 
providing a nonvanishing contribution at $|{\bf k}|>k_F$. 

\begin{figure}[hbt]
\centerline
{\psfig{file=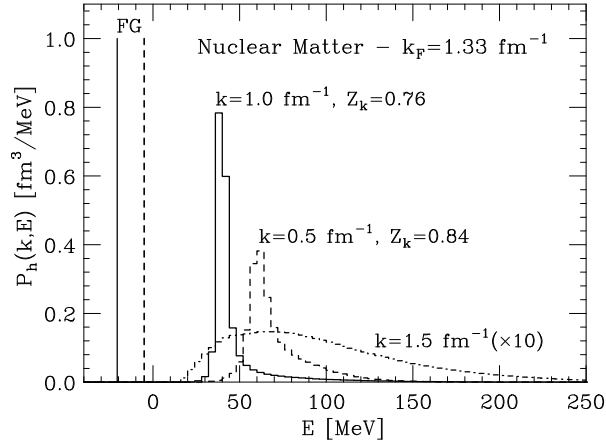,width=8cm}}
\caption{ Energy dependence of the hole spectral function of nuclear
matter at equilibrium density, corresponding to $k_F = 1.33$ fm$^{-1}$. 
The solid, dashed and dot-dash lines
correspond to $|{\bf k}|=$ 1, 0.5 and 1.5 fm$^{-1}$,
respectively. The FG spectral function at $|{\bf k}|=$ 1 and 0.5 fm$^{-1}$ is shown
for comparison.  \label{figpke} }
\end{figure}

Figure \ref{figpke} illustrates the energy dependence of the hole spectral function
of nuclear matter, calculated in Ref.\cite{pke} using the correlated basis approach. 
Comparison with the FG model clearly shows
that the effects of nuclear dynamics and NN correlations are large, resulting
in a shift of the quasiparticle peaks, whose finite width becomes
large for deeply-bound states with $|{\bf k}| \ll k_F$. In addition, NN
correlations are responsible for the appearance of strength at $|{\bf k}|>k_F$.

The results of nuclear matter calculations have
been extensively employed to obtain the hole spectral functions of heavy nuclei 
within the local density approximation (LDA) \cite{LDA}.

\section{Nuclear response to a scalar probe}
\label{response}

Within NMBT, the nuclear response to a scalar probe delivering
momentum {\bf q} and energy $\omega$ can be written in terms of the the
imaginary part of the particle-hole propagator $\Pi({\bf q},\omega)$ according
to \cite{FetterWalecka,BFF:1992}
\beq
\label{def:resp}
S({\bf q},\omega) = \frac{1}{\pi}\ {\rm Im}\  \Pi({\bf q},\omega) = \frac{1}{\pi}\
{\rm Im}\ \langle 0 \vert \rho^\dagger_{{\bf q}}  \
\frac{1}{H-E_0-\omega-i\eta} \ \rho_{{\bf q}} \vert 0 \rangle \ ,
\eeq
where $\eta=0^+$, $\rho_{{\bf q}}= \sum_{{\bf k}} a^\dagger_{{\bf k}+{\bf q}} a_{{\bf k}}$
is the operator describing the fluctuation of the target density induced by the 
interaction
with the probe,
$a^\dagger_{{\bf k}}$ and $a_{{\bf k}}$ are nucleon creation
and annihilation operators, and $\vert 0 \rangle$ is the
target ground state, satisfying the Schr\"odinger equation
$H\vert 0 \rangle = E_0 \vert 0 \rangle$.

In general, the calculation of the response requires the knowledge of the spectral 
functions associated with both particle and hole states, as 
well as of the particle-hole effective interaction \cite{BFF:1992,WIM:2004}.
The spectral functions are mostly affected by short range NN correlations
(see Fig. \ref{figpke}), while the inclusion of the effective interaction, e.g. within the
framework of the Tamm Dancoff and Random Phase Approximation \cite{WIM:2004,benhar09}, 
is needed to account
for collective excitations induced by long range correlations, involving more than
two nucleons.

At large momentum transfer, as the space resolution of the probe becomes small compared
to the average NN separation distance, $S({\bf q},\omega)$ is no longer significantly
affected by long range correlations \cite{benhar09}. In this kinematical regime the zero-th order
approximation in the effective interaction, is expected to be applicable. The response 
reduces to the incoherent sum of contributions coming from
scattering processes involving a single nucleon, and can be written in the simple form
\beq
\label{L0}
S({\bf q},\omega) = \int d^3k dE\ P_h({\bf k},E) P_p({\bf k}+{\bf q},\omega-E) \ .
\eeq
The widely employed impulse approximation (IA) can be readily obtained from
the above definition replacing $P_p$ with the prediction of the FG model, which amounts to
disregarding final state interactions (FSI) betwen the struck nucleon and the spectator
particles. The resulting expression reads
\beq
\label{IA}
S_{IA}({\bf q},\omega) = \int d^3k dE\ P_h({\bf k},E) \theta(|{\bf k}+{\bf q}|-k_F)
\delta(\omega-E-\epsilon^0_{|{\bf k}+{\bf q}|}) \ .
\eeq

\begin{figure}[hbt]
\centerline
{\psfig{file=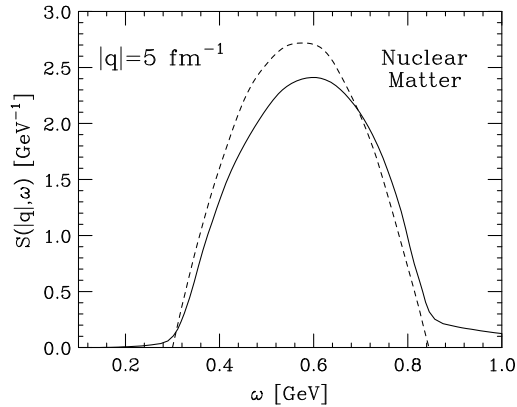,width=6.8cm}}
\caption{ Nuclear matter $S_{IA}({\bf q},\omega)$ (see Eq.(\ref{IA})), as a function
of $\omega$ at $|{\bf q}|=5$ fm$^{-1}$.  The solid and dashed lines correspond to
the spectral function of Ref.\protect\cite{pke} and to the FG model, 
respectively. \label{iscorr} }
\end{figure}

Figure \ref{iscorr}, showing the $\omega$ dependence of the nuclear matter response
function at $|{\bf q}|=5$ fm$^{-1}$, illustrates the role of correlations
in the target initial state. The solid and dashed lines have been obtained from
Eq.(\ref{IA}), using the spectral function of Ref.\cite{pke}, and the
from the FG model,  
 respectively. It is apparent that the inclusion of correlations produces a
significant shift of the strength towards larger values of energy transfer.

Obvioulsy, at large ${\bf q}$ the calculation of $P_p({\bf k}+{\bf q},\omega-E)$ cannot
be carried out using a nuclear potential model. Hovever, it can be
obtained form the measured NN scattering amplitude within the eikonal
approximation. A systematic scheme to include corrections to Eq.(\ref{IA}) and take into
account FSI has been developed in Ref.\cite{gangofsix}. 
The main effects of FSI on the response are i) a
shift in energy, due to the mean field of the spectator nucleons and ii) a
redistributions of the strength, due to the coupling of the one particle-one hole
final state to $n$ particle-$n$ hole final states.

\begin{figure}[hbt]
\centerline
{\psfig{file=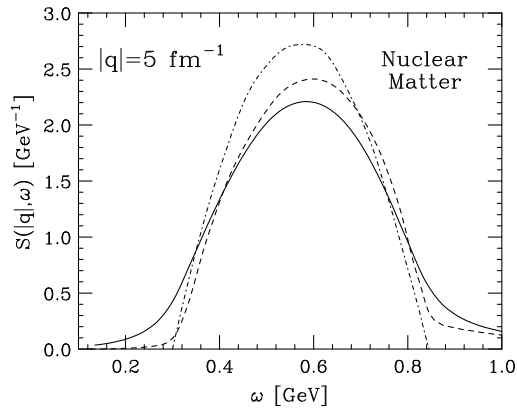,width=6.8cm}}
\caption{Nuclear matter $S({\bf q},\omega)$ as a function
of $\omega$ at $|{\bf q}|=5$ fm$^{-1}$.  The solid and dashed lines have been obtained
from the spectral function of Ref. \cite{pke}, with and without inclusion
of FSI, respectively. The dot-dash line corresponds to the FG model. \label{fsicorr} }
\end{figure}

Figure \ref{fsicorr} shows the $\omega$ dependence of the nuclear matter response of
Eqs.(\ref{L0}) and (\ref{IA}) at $|{\bf q}|=5$ fm$^{-1}$. 
The solid and dashed lines have been
obtained using the spectral function of Ref.\cite{pke}, with and without inclusion
of FSI according to the formalism of Ref.\cite{gangofsix}, respectively.
For reference, the results of the FG model are also shown by the dot-dash line.
The two effects of FSI, energy shift and redistribution of the strength from the region of
the peak to the tails, clearly show up in the comparison betweem soild and dashed lines.

\section{Electron-nucleus cross section}
\label{eA}

The differential cross section of the process
\beq
e + A \rightarrow e^\prime + X \ ,
\label{eA:process}
\eeq
in which an electron of initial four-momentum $k_e\equiv(E_e,{\bf k}_e)$ scatters off
a nuclear target to a state of four-momentum
$k^\prime_e\equiv(E_{e^\prime},{\bf k}_{e^\prime})$, the target final state
being undetected, can be written in Born approximation as
\beq
\frac{d^2\sigma}{d\Omega_{e^\prime} dE_{e^\prime}} =
\frac{\alpha^2}{Q^4}\frac{E_{e^\prime}}{E_e}\ \Lmunu\Wmunu \ ,
\label{eA:xsec}
\eeq
where $\alpha=1/137$ is the fine structure constant,
$d\Omega_{e^\prime}$ is the differential solid angle in the direction
specified by ${\bf k}_{e^\prime}$, $Q^2=-q^2$ and
$q=k_e-k_{e^\prime} \equiv (\omega,{\bf q})$ is the four momentum transfer.

The tensor $\Lmunu$ is fully specified by the measured
electron kinematical variables. All the information on target structure is contained
in the tensor $\Wmunu$, whose definition involves the initial and final nuclear
states $|0\rangle$ and $|X\rangle$, carrying four-momenta $p_0$ and $p_X$,
as well as the nuclear current operator $J^\mu$:
\beq
\Wmunu=\sum_{X}  \langle 0 |J^\mu|X \rangle\langle X|J^\nu|0\rangle
\delta^{(4)}(p_0+q-p_X)\ ,
\label{nuclear:tensor}
\eeq
\noindent
where the sum includes all hadronic final states. Note that the tensor
of Eq.(\ref{nuclear:tensor}) is the generalization of the nuclear response,
discussed in the previous section, to the case of a probe interacting
with the target through a vector current. To see this, insert the complete set 
of eigenstates of the nuclear hamiltonian 
 in the definition of Eq.(\ref{def:resp}). The result is
\beq
S({\bf q},\omega) = \sum_n  \langle 0 | \rho^\dagger_{{\bf q}} | n \rangle
\langle n | \rho_{{\bf q}} | 0 \rangle \delta(\omega + E_0 - E_n ) \ ,
\eeq
to be compared to Eq.(\ref{nuclear:tensor}).

In the IA regime, the nuclear
current appearing in Eq.~(\ref{nuclear:tensor}) can be written as a sum of one-body
currents
\beq
J^\mu \rightarrow \sum_i j_i^\mu \ ,
\label{currIA}
\eeq
while $|X\rangle$ reduces to the direct product of the hadronic state produced at the
electromagnetic vertex, carrying four momentum $p_x \equiv(E_x,{\bf p}_x)$, and the state 
describing the residual system, carrying momentum
${\bf p}_{\cal R}= {\bf q}-{\bf p}_x$.

As a result, the Eq.~(\ref{nuclear:tensor}) can be rewritten in the form
($k\equiv(E,{\bf k})$)
\beq
\Wmunu({\bf q},\omega) = \int d^4k \ 
\left(\frac{m}{E_{{\bf k}}}\right) \left[ Z P_p(k)
w_p^{\mu\nu}({\widetilde q}) +  N P_n(k)
w_n^{\mu\nu}({\widetilde q}) \right] \ ,
\label{hadrten2}
\eeq
where $Z$ and $N=A-Z$ are the number of target protons and neutrons, while
$P_p$ and $P_n$ denote the proton and neutron {\em hole} spectral functions,
respectively. In Eq.~(\ref{hadrten2}), $E_{{\bf k}} = \sqrt{|{\bf k}^2|+m^2}$ and
\beq
w_N^{\mu\nu} = \sum_x \langle {\bf k},{\rm N}| j^\mu_N | x,{\bf k}+{\bf q} \rangle
\langle {\bf k}+{\bf q},x | j^\nu_N | {\rm N},{\bf k} \rangle
 \delta({\widetilde \omega} + E_{{\bf k}} - E_x ) \  .
\label{nucleon:tensor}
\eeq
The tensor $w^{\mu\nu}_n$ describes the electromagnetic structure 
  of a nucleon of initial momentum ${\bf k}$
{\em in free space}. The effect of nuclear binding is accounted for by the
replacement $\omega \rightarrow {\widetilde \omega}$, with \cite{RMP}
\beq
{\widetilde \omega} = E_x - E_{{\bf k}} = 
\omega - E + m - E_{{\bf k}} \ .
\label{omega:tilde}
\eeq

The above equations show that within the IA scheme, the definition of the
electron-nucleus cross section involves two elements: i) the tensor
$w_N^{\mu\nu}$, that can be extracted from electron-proton and electron-deuteron data,
and  ii) the spectral function, discussed in the Section \ref{NMBT}.

The formalism of NMBT has been extensively employed in the analysis of a variety 
of electron-nucleus scattering observables. In Ref. \cite{Benhar05}, it has been employed
to calculate the inclusive electron scattering cross sections off oxygen, at beam
energies ranging between 700 and 1200 MeV and electron scattering angle 32$^\circ$.
In this kinematical region single nucleon
knock out is the dominant reaction mechanism and both quasi-elastic
and inelastic processes, leading to the appearance of nucleon resonances,
must be taken into account.

\begin{figure}[hbt]
{\psfig{figure=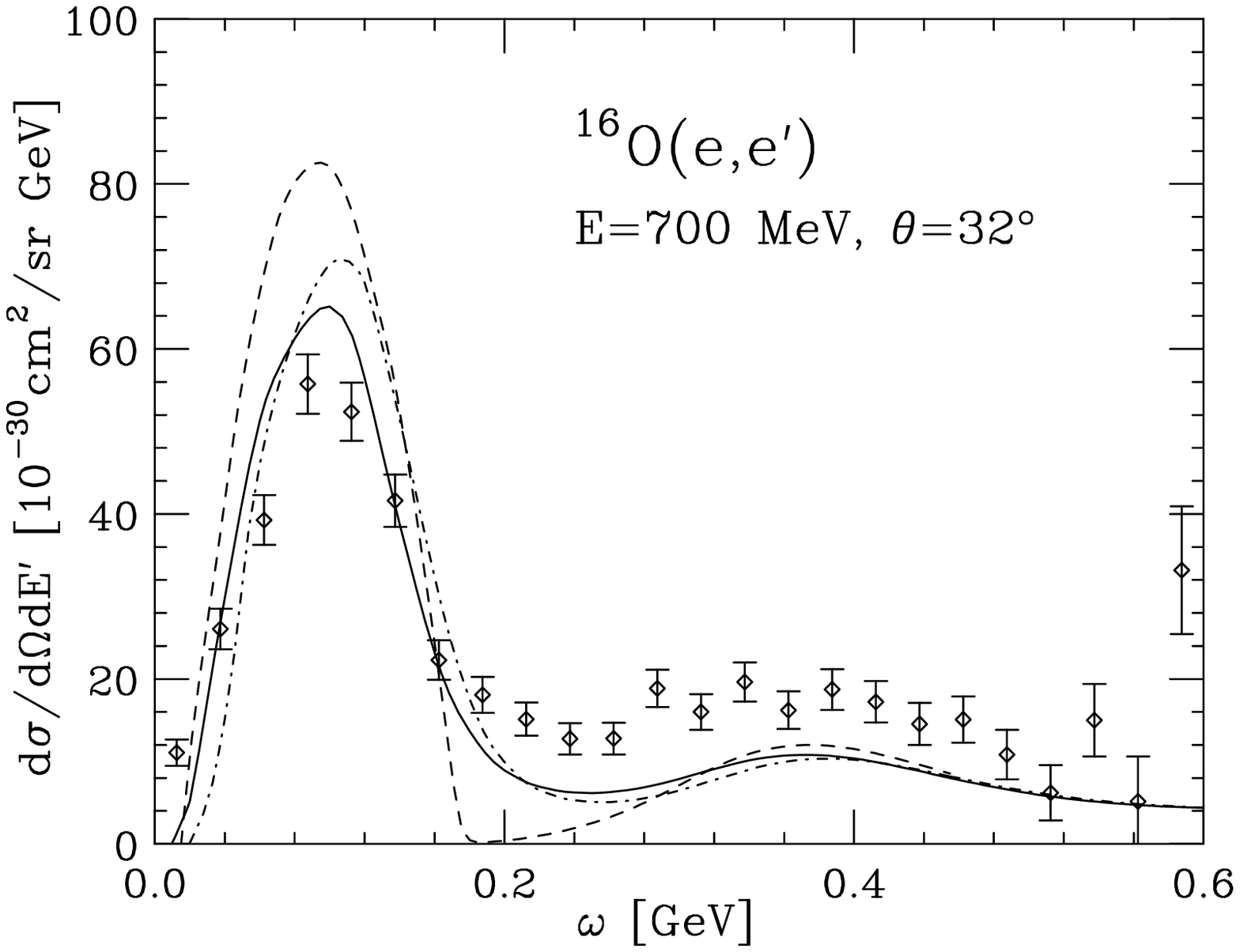,angle=00,width=6.2cm,height=5.0cm}}
{\psfig{figure=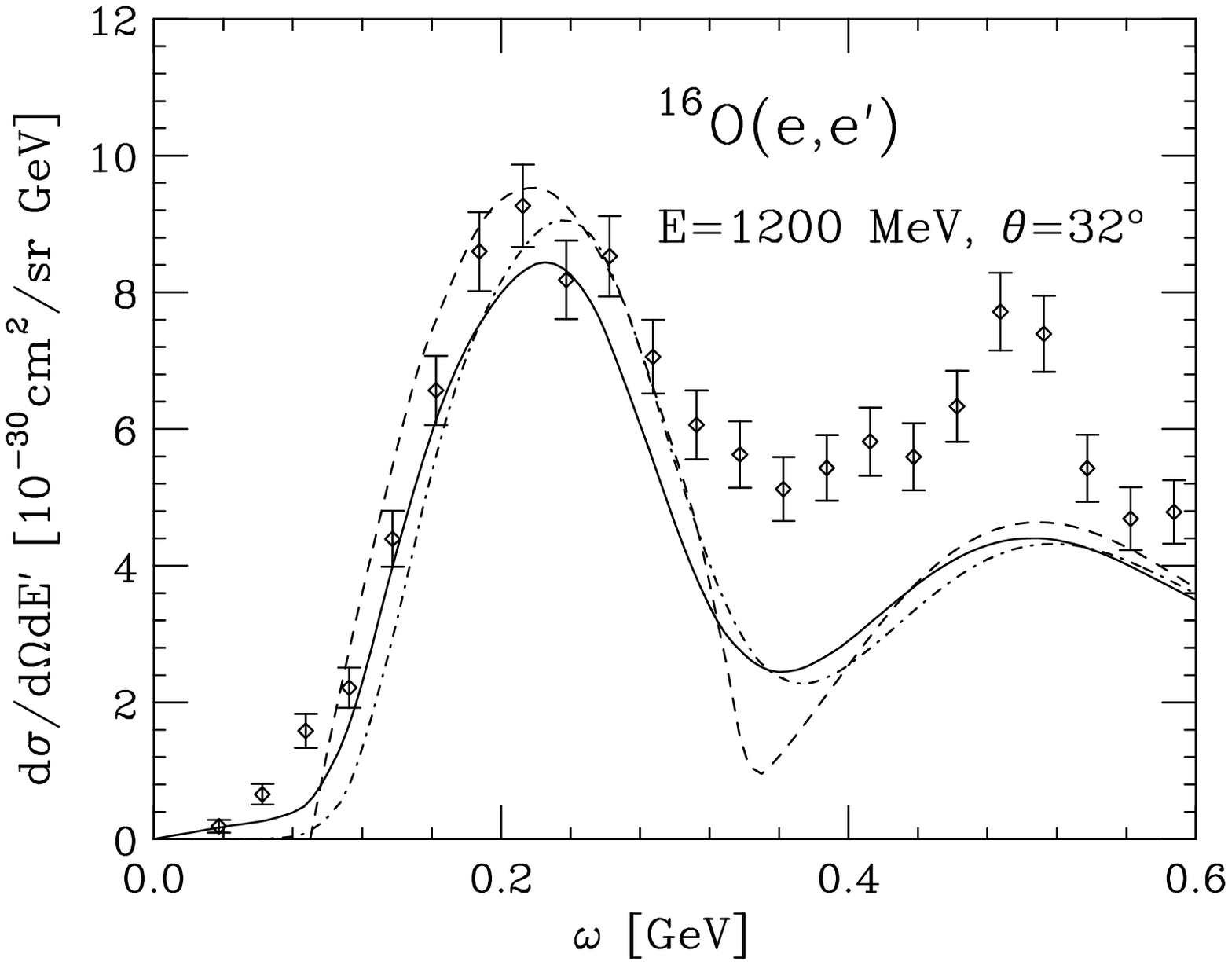,angle=00,width=6.2cm,height=5.0cm}}
\caption{ Cross section of the process $^{16}O(e,e^\prime)$
at scattering angle 32$^\circ$ and beam energy 700 MeV (left panel) and 1200 MeV
(right panel), as a function of the electron energy loss $\omega$.
Solid lines: full calculation, including FSI. Dot-dash lines: IA calculation.
Dashed lines: FG model. The data are taken
from Ref.\protect\cite{LNF} \label{fig:ee}}
\end{figure}

The comparison between theory and the experiment, in Fig. \ref{fig:ee}, shows 
 that the data in the region of the quasi-elastic peak are accounted for
with an accuracy better than $\sim$ 10 \%. The discrepancies observed at 
larger electron energy loss, where $\Delta$ production
dominates, can be ascribed to deficiencies in the description of the nucleon structure 
functions \cite{Delta}. For reference, the predictions of the FG model are also displayed 
by dashed lines. A realistic description of nuclear dynamics
clearly appears to be needed to explain the measured cross sections.

\section{Charged current neutrino-nucleus cross section}
\label{nuA}

The cross section of the weak charged current process $\nu_\ell + A \rightarrow \ell^- + X$ 
can be written in the form (compare to Eq. (\ref{eA:xsec}))
\beq
\frac{d^2\sigma}{d\Omega_\ell dE_\ell}=\frac{G_F^2\,V^2_{ud}}{16\,\pi^2}\,
\frac{|\bf k_\ell|}{|\bf k|}\,L_{\mu\nu}\, W_A^{\mu\nu} \ ,
\label{nu:cross:section}
\eeq
where $G_F$ is the Fermi constant, $V_{ud}$ is the CKM matrix element
coupling $u$ and $d$ quarks and ${\bf k}$ and ${\bf k}_\ell$ denote the momenta of the 
incoming neutrino and the outgoing charged lepton, respectively. 

The formalism outlined in the previous section can be readily 
generalized to the case of neutrino-nucleus interactions, the required nuclear physics
input being the same in the two instances. On the other hand, while the vector form 
factors entering 
the definition of the electron-nucleus cross section can be measured 
with great accuracy using proton and deuteron targets, the experimental 
determination of the nucleon axial form factor is still somewhat controversial, 
as different experiments report appreciably different results \cite{oldmass,K2K,BOONE,nomad}.
In these lectures, I will focus on the role of nuclear dynamics, and do not discuss the 
uncertainty associated with the weak form factor.

\begin{figure}[hbt]
\centerline%
{\psfig{figure=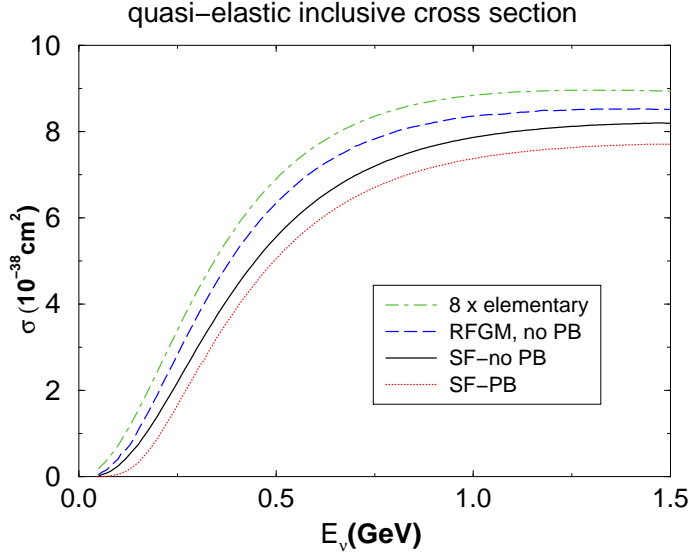,angle=00,width=9.0cm}}
\caption{\small Total quasi-elastic cross section of the process
$\nu_e + ^{16}O \rightarrow e^- +  X$.
The dot-dash line represents eight times the elementary cross section; the dashed
line is the result of the FG model; the
dotted and solid lines have been obtained using the spectral function of
Ref. \cite{LDA}, with and without inclusion of Pauli blocking, respectively.
 \label{nu_QE1}}
\end{figure}

In order to gauge the magnitude of nuclear effects, in 
Fig. \ref{nu_QE1} the energy dependence of the quasi elastic contribution
to the total cross section of the process $\nu_e + ^{16}O \rightarrow  e^- + X$
computed using different approximations are compared \cite{benhar07}. 
The dot-dash line represents the
result obtained describing oxygen as a
collection of noninteracting stationary nucleons, while the dashed and
solid lines have been obtained from the FG model and using 
 the spectral function of Ref. \cite{LDA}, respectively.
It is apparent that replacing the FG with the approach based on
a realistic spectral function leads to a sizable suppression of the total
cross section. Comparison between the dot-dash line and the dotted one,
obtained taking into account the effect of Pauli blocking \cite{Benhar05}, 
shows that the overall change due to nuclear effect is $\sim$~20~\%.
                                                                                  
Note that FSI between the nucleon produced at the elementary weak interaction
vertex and the spectator particles have not been taken into account, as they
{\em do not} contribute to the total cross section.

To see how much the description of nuclear dynamics may affect the data analysis 
of neutrino oscillation experiments, consider reconstruction of the incoming 
neutrino energy in charged current quasi elastic events 
$\nu_\mu + A \rightarrow \mu + p + (A-1)$, in which the muon energy, $E_\mu$, and 
angle, $\theta_\mu$, are measured.

From the requirement that the elementary scattering process be elastic, 
it follows that the neutrino energy is given by
\beq
E_\nu=\frac{m_p^2-m_\mu^2-E_n^2+2E_\mu E_n- 2{\bf k}_\mu \cdot {\bf p}_n+|{\bf p}_n^2|}
{2( E_n - E_\mu + |{\bf k}_\mu|\cos \theta_\mu - |{\bf p}_n|\cos \theta_n)} \ ,
\label{kin1}
\eeq
where $m_p$ and $m_\mu$ denote the proton and muon mass, respectively, 
${\bf k}_\mu$ is the muon momentum and ${\bf p}_n$ and $E_n$ are the momentum 
and energy carried by the struck neutron.

Setting $|{\bf p}_n| = 0$ and fixing the neutron removal energy to
a constant value $\epsilon$, i.e. setting $E_n = m_n - \epsilon$, $m_n$ being the 
neutron mass, Eq.(\ref{kin1}) reduces to
\beq
E_\nu = \frac{2E_\mu(m_n - \epsilon)-
(\epsilon^2 - 2 m_n \epsilon + m_\mu^2 + \Delta m^2) }
{2 ( m_n - \epsilon - E_\mu + |{\bf k}_\mu|\cos \theta_\mu )} \ ,
\label{simple:kin}
\eeq
with $\Delta m^2 = m_n^2 - m_p^2$. In the analysis of Refs. \cite{K2K,BOONE}
the energy of the incoming neutrino has been reconstructed using the above
equation.

\begin{figure}[hbt]
\centerline%
{\psfig{figure=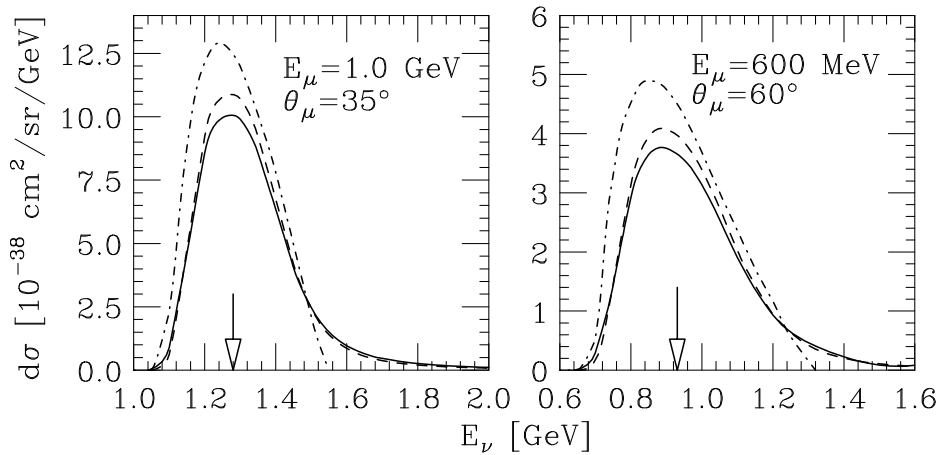,angle=00,width=12.5cm}}
\caption{\small 
Right panel: Differential cross section of the process
$\nu_\mu + A \rightarrow \mu + p + (A-1)$, at $E_\mu=$ 600 MeV
and $\theta_\mu=$ 60$^\circ$, as a function of the incoming neutrino energy.
The solid line shows the results of the
full calculation, carried out within the approach of Refs. \cite{Benhar05,benhar07},
whereas the dashed line has been obtained neglecting the effects
of FSI. The dot-dash line corresponds to the
 FG model. The arrow points to
the value of $E_\nu$ obtained from Eq. (\ref{simple:kin}).
Left panel: Same as the right panel, but for $E_\mu=$ 1 GeV and
$\theta_\mu=$ 35$^\circ$. \label{nuspec1} }
\end{figure}

The differences between the $E_\nu$ predicted by the approach based on a 
realistic spectral function and that obtained from the FG model is illustrated 
in Fig. (\ref{nuspec1}), where the values obtained from Eq. (\ref{simple:kin}) are 
also shown by arrows. The appearance of the tail extending to large $E_\nu$, 
to be ascribed to NN correlations not included in the FG model, leads to a 
sizable increase of the average neutrino energy. 

\section{Conclusions}
\label{concl}

Dynamical correlation effects, which are long known to play a critical role 
in shaping the nuclear response to electromagnetic probes, are also 
important in neutrino-nucleus interactions.

Although the answer to the question addressed in the title of these lectures 
is somewhat context dependent, as not all the observables measured in neutrino
experiments are equally sensitive to NN correlations, there 
are instances in which a realistic description of nuclear structure and dynamics 
is badly needed. For example, analyses aimed at extracting 
{\em nucleon} properties, such as the axial form factor, from {\em nuclear}
cross sections require a fully quantitative control of nuclear effects.

The formalism based on NMBT, which proved very effective in theoretical studies 
of electron-nucleus scattering, can be easily generalized to the case of weak 
interactions. The implementation of realistic spectral functions in the Monte Carlo
simulation codes, which would significantly improve the description of the initial 
state, does not involve severe difficulties. As far as final states
are concerned, a consistent description of FSI effects is available for the case
of quasielastic scattering, which is the dominant reaction mechanism at beam 
energies around 1 GeV. The extension to the case of pion production and deep 
inelastic scattering is certainly possible, and is being actively investigated.

\end{document}